\documentclass[preprint,showpacs,preprintnumbers,amsmath,amssymb,nofootinbib]{revtex4}
\usepackage{epsfig}
\usepackage{graphicx}
\usepackage{dcolumn}
\usepackage{bm}
\usepackage{threeparttable}
\usepackage{lipsum}  
 \usepackage{float}

\newcommand{\beq}{\begin{equation}}
\newcommand{\eeq}{\end{equation}}
\newcommand{\bea}{\begin{eqnarray}}
\newcommand{\eea}{\end{eqnarray}}
\newcommand{\ba}{\begin{array}}
\newcommand{\ea}{\end{array}}
\newcommand{\bi}{\bibitem}

\newcommand{\Ht}{\widetilde H}
\newcommand{\Lt}{\widetilde L}
\newcommand{\Z}{\widehat R}

\def\pr{\prime}
\let\jnfont=\rm

\def\PLB{\jnfont Phys.\ Lett.\ B}

\def\PRD{\jnfont Phys.\ Rev.\ D}
\def\PRL {\jnfont Phys.\ Rev.\ Lett.}

\def\PLB{\jnfont Phys. Lett. B}

\newcommand{\nm}{\nonumber}

\begin{document}
\title{Leptonic Flavor Changing Processes $\ell_i \to \ell_j\gamma$ and $\ell_i \to \ell_j\ell_k\ell_l$ in the 
Twin Higgs Models}

\author{Guo-Li Liu$^{1}$,\footnote{guoliliu@zzu.edu.cn}, Fei Wang$^{1}$,\footnote{feiwang@zzu.edu.cn}, Wenyu Wang$^2$,\footnote{wywang@bjtu.edu.cn}}
\affiliation{ School of Physics and Microelectronics, Zhengzhou University, Zhengzhou, 450001, China \\
 $^2$ Institute of Theoretical Physics, Faculty of Science, Beijing University of Technology, Beijing 100124, China}

\begin{abstract}
Heavy neutrinos are usually introduced
to accommodate tiny neutrino masses via seesaw mechanism,
or to alleviate the cosmology problem,
and there may exist charged Higgses which couple to the leptons
with different flavors.
 These two features can both appear in the 
 Twin Higgs models.
What interests us is that such new particles and interactions may lead to new contributions to the lepton flavor violating
 processes $\ell_i \to \ell_j\gamma$ and $\ell_i \to \ell_j\ell_k\ell_l$.
We find that current experimental data can 
constrain the parameter spaces and certain lepton flavor violating processes can possibly be tested by the next generation experiments.
%
\end{abstract}

\maketitle
\section{Introduction}

In the Standard Model (SM), the conservation of individual lepton flavor numbers
and the presentation of the GIM mechanism~\cite{GIM-1970} guarantee
that the cross sections and/or branching ratios of the lepton flavor violating
(LFV) processes are tiny. 
For example, the SM predicts the branching ratio $BR_{SM}(\mu\to e\gamma)\sim 10^{-55}$~\cite{SMljlig}
to be very small, which would never be probed in experiments.
Therefore, it will be an obvious evidence of new physics
beyond the SM if we observe any signature of the LFV processes
in future experiments, so LFV processes are very efficient in exploring new physics beyond the SM.
As the upper bounds of the LFV processes by current experiments
have been given with impressive accuracy, it is interesting to explore
whether theoretical predictions of a typical new physics model can agree with
the experimental results.
 Many works have been completed to look for the
  LFV processes $\ell_i \to \ell_j\gamma$, $\ell_i \to \ell_j\ell_k\ell_l$
in new physics models, see e.g. ~\cite{bSM1,bSM2,bSM3,bSM4,bSM5,bSM6,
1701.00947,1812.03860,1812.02449,1912.05900,0506070,0307126,0206056}.

It is well known that the neutrino oscillation experiments
~\cite{neutrino1,neutrino2,neutrino3,neutrino4,neutrino5,neutrino6,neutrino7,neutrino8,neutrino9}
 indicate that neutrinos possess tiny masses and can convert into each other,
so individual lepton flavor numbers $L_i=L_e,~L_\mu,~L_\tau$ may be violated at the electroweak scale.
Weinberg's effective dimension-5 operator is the lowest one that can generate tiny Majorana-type neutrino masses.
Such an operator can be tree level ultraviolet(UV)-completed to obtain three types of tree-level seesaw mechanism:
type I seesaw~\cite{type I}, involving the exchange of right-handed neutrinos;
 type II seesaw~\cite{type II}, involving the exchange of scalar triplet;
 type III~\cite{type III}, involving the exchange of fermion triplet.
 The neutrino seesaw mechanisms can also be embedded into various well
 motivated new physics models to accommodate the neutrino masses,
 such as the 
 Twin Higgs (TH) models~\cite{litt-hier1}.

TH models, which usually extend the SM with a copy, can stabilize the Higgs mass from quadratic divergent quantum corrections by a discrete 
 symmetry.
 The new particles, which are related to the SM particles by this discrete symmetry, do not carry the SM color charges and can hardly be produced and detected at the colliders, ameliorating the stringent constraints on the mass of
 the top partner by the Large Hadron Collider(LHC)~\cite{parti-tev}. 

 Although TH models are theoretically appealing, they encounter difficulties by cosmological considerations.
In the simplest realization of TH ~\cite{litt-hier1} models,
the twin particles will eventually transfer their entropies into twin photons and twin neutrinos,
which behave as extra radiation components, increasing the SM prediction by an additional amount  $\Delta N_{eff} \sim 5.6$ ~\cite{1611.07975-neff_mth}, which deviates the observations~\cite{neff_exp1,neff_exp2},  where $N_{eff} $ is effective number of (light) neutrino species.

Various modifications are proposed to reduce the $N_{eff}$ value and
  solve the problems of TH models.
For example, the first and second generations of twin fermions as well as twin photon
  can be absent~\cite{modif-mth1};
Or in some realizations, the twin neutrinos, and even twin photons can be heavy
~\cite{modif-mth2,modif-mth3-1703.06884,modif-mth4,modif-mth5-1905.00861}.
  Asymmetric entropy production can be possible after the twin and
  SM sectors decouple~\cite{1611.07975-neff_mth,modif-mth3-1703.06884,modif-mth6,modif-mth7}.

The seesaw extension of TH models usually predicts massive twin neutrino masses
to lower the effective degrees of freedom contributed by the twin sector~\cite{modif-mth3-1703.06884}.
These models can also introduce extra massive right-handed neutrinos
in both sectors of SM and the twin~\cite{1611.07975-neff_mth}
to generate the tiny neutrino masses.
Typical lepton-flavor-changing couplings will appear via these mechanisms,
which may lead to interesting phenomenological consequences,
so we try to study the lepton flavor violating (LFV) processes
 $\ell_i\to \ell_j\gamma$, $\ell_i\to \ell_j \ell_k \ell_l$ induced by these new couplings
in this work.

The paper is organized as follows.
Section II briefly gives the new LFV couplings in TH models. 
In section III, the amplitudes and the branching ratios of rare
LFV processes $\ell_i \to \ell_j\gamma$, $\ell_i\to \ell_j \ell_k \ell_l$ will be calculated, respectively.
The corresponding numerical results are also shown in this section.
The conclusion will be given in Section IV.

\section{The Lepton Flavor Violating Couplings in the twin higgs models}
In TH framework, the SM particles and their 
copies are related by the discrete $Z_2$ twin symmetry.
To contain a residual custodial symmetry,
the global symmetry of the Higgs sector in the simplest realization
can be taken as $SO(8)$ or $SO(7)$ \cite{Barbieri:2015lqa,Batra:2008jy,1501.07890,1905.02203}.
The SM Higgs doublet is a part of the pseudo-Nambu-Goldstone bosons (pNGBs),
which arises from the spontaneously breaking of the global $SO(8)$($SO(7)$ ) symmetry into $SO(7)$( $G_2$).
The neutral Higgs mass, under the joint action of the global symmetry and the discrete twin symmetry,
is protected from one loop quadratic divergence.


Besides the SM-like neutral Higgs, there would be charged Higgses.
In some TH models, for different goals, extra scalars will be introduced.
For example, in Ref.\cite{1702.04399},
a $SU(2)_L$ singlet charged scalar $S^+$ is introduced to provide suitable
neutrino masses via couplings to the right-handed neutrinos,
while in Refs.\cite{modif-mth4, 
modif-mth5-1905.00861}, a new scalar $\phi$ is added to have
the similar effect in the couplings with the leptons.
Another possiblity for the extra charged scalars to appear in the particle list is
due to the enlarging breaking mode.
To keep the breaking smallest, the economical breaking choices are chosen as above
but it can be otherwise in some situation.
For example, the breaking modes can be $SO(8)\to G_2$ or $SO(2N)\to SO(N)\times SO(N)$ \cite{2202.01228}
(the former $N$ is for SM sector, while the latter, for twin sector).




After the global breaking such as $SO(8)\to G_2$, the seven pNGBs
can be parameterized via the decomposition $\mathbf{8} = (\mathbf{2},\mathbf{1},\mathbf{2}) + (\mathbf{1},\mathbf{2},\mathbf{2})$ under $SU(2)_L \times SU(2)_{\Lt} \times SU(2)_{\Z}$ as
\beq
\label{higgses}
(\mathbf{2},\mathbf{1},\mathbf{2}): \, H
= {f \over \sqrt{2}} \begin{pmatrix} \pi_2 + i \pi_1 \\ h - i \pi_3 \end{pmatrix} \, , \quad
(\mathbf{1},\mathbf{2},\mathbf{2}): \, \Ht
= {f \over \sqrt{2}} \begin{pmatrix} \pi_6 + i \pi_5 \\ \sigma - i \pi_7 \end{pmatrix} \, ,
\eeq
where $f$ is the Goldstone scale, $\sim$ TeV, and $\sigma = \sqrt{1-\pi_{\hat a}^2}$ ($\hat a=1,...7$),
and $h \equiv \pi_4$ is assumed to be the SM-like Higgs.
The SM gauge fields acquire the typical masses proportional to the scale of electroweak
symmetry breaking, $v$.

In case additional contributions to $N_{eff}$ of TH models cause difficulties in cosmology\footnote{In scenarios with multiple dark matter components, small contributions to $N_{eff}$ can possibly solve the discrepancy between the value of $H_0$ extracted from local measurements versus CMB data.},
one can impose another portal, such as the neutrino portal to reduce the
$N_{eff}$ to the acceptable level. However, new flavor-changing couplings can be generated  via the mixing between the SM neutrino and the twin neutrino and they can be given as~\cite{modif-mth3-1703.06884},
\bea
\mathcal{L_{{\rm int}(\nu,\tilde\nu)}} &=& \frac{g}{\sqrt{2}}\bar{\ell} \gamma^\mu P_L (c_\theta\nu + s_\theta \tilde{\nu})  W^+_\mu + {\rm h.c.} \nm\\
&=& V_{\tilde{\nu}} \bar{\ell} \gamma^\mu P_L \tilde{\nu}  W^+_\mu + {\rm h.c.} +...
\label{lfv-1}
\eea
where $c_\theta = \cos\theta$,  $s_\theta = \sin\theta$ and $\theta$ is the  mixing angle between ordinary neutrino $\nu$ and twin neutrino $\tilde{\nu}$.
Note that there is actually not only the  mixing,
but also the Pontecorvo-Maki-Nakagawa-Sakata (PMNS)~\cite{mns-maki-1962,0712.4019} matrix elements.
Here for simplicity we denote $V_{\tilde{\nu}}$ to represent the effect of all these parameters.
So does the following parameter $y_{\nu_R}$.

On the other hand, the LFV couplings can also be induced from the typical neutrino seesaw extension
of the TH models via introducing the massive right-handed neutrino~\cite{1611.07975-neff_mth,1905.02203}, 
\bea \nm
{\cal L}&\supset & -y_{ij} (L^i_A H_A N^j_A + L^i_B H_B N^j_B)-\frac{1}{2}(M_N)_{ij}(N_A^iN_A^j+N_B^iN_B^j)-(M_{AB})_{ij}N_A^iN_B^j~ +h.c.\\
&=& - y_{\nu_R} \bar{\ell}  \nu_R  H^+ + {\rm h.c.} +...,
\label{lfv-2}
\eea
in both the SM(denoted by $A$) and twin sectors(denoted by $B$).

From Eqs. (\ref{lfv-1}) (\ref{lfv-2}), we can see the particles
which induce the LFV couplings could be gauge bosons
and the new charged scalars $H^\pm$ with the twin neutrinos $\tilde{\nu}$ or the right-handed neutrinos $\nu_R$.
Note that there are two kinds of neutrinos:
In general, the twin neutrinos $\tilde{\nu}$ can couple to gauge bosons together with the leptons,
and the right-handed neutrinos $\nu_R$ couple
to the charged Higgs with the leptons.
From the following discussion,
we can see that they both are much heavier than the ordinary neutrinos,
so we will call them "heavy neutrinos" when not causing confusion.

In the process of lowering the decoupling temperature and potentially reducing the contribution to
$N_{eff}$, the constraint on the mixing of the ordinary neutrino $\nu$ and twin neutrino $\tilde{\nu}$ is mainly from
the decay of the twin neutrino $\tilde{\nu}$ to the ordinary neutrino $\nu$,  $\tilde{\nu}\to \nu f\bar f$
and the  semi annihilations $\tilde{\nu} \nu \to \nu\nu$, with the usual range scope of
the value of the decoupling temperature, $0.3~GeV<T_d<1~GeV $ \cite{modif-mth3-1703.06884,1604.02458},
and the favourable space can be ${\bf10^{-4}<V_{\tilde{\nu}} < 10^{-3}}$ with twin neutrino mass of $10$ GeV,
which is the optimal value for the thermal decoupling between the SM and the twin sectors.

Normally, to provide the tiny neutrino masses, we need massive right-handed neutrinos,
as those in type I seesaw.
The charged Higgs mass is discussed in references such as \cite{0611015-su,lfv-lrth,charged-neutral-mass} and our previous works\cite{works-inv-chaghiggs}.
According to the references, the mass difference between the charged Higgs boson and the neutral heavy Higgs boson
should be less than $300$ GeV, i.e., $|m_{H^\pm}- m_0|\leq 300$ GeV, and in general, the mass of the
neutral heavy Higgs boson is larger than that of the SM-like Higgs.
In our discussion and calculation we can take $m_H^\pm$ and $m_{\nu_R}$ in the following ranges
\beq
100 ~{\rm GeV} <m_{H^\pm}< 1000~ {\rm GeV},~~500~GeV<m_{\nu_R}<5000~GeV.
\eeq

As for the Yukawa couplings $y_{\nu_R}$, according to the relationship of the masses of
the ordinary neutrino and the right-handed neutrino, $m_\nu \sim \frac{y_{\nu_R}^2v^2}{m_{\nu_R}}$~\cite{1611.07975-neff_mth},
we can estimate the $y_{\nu_R}$ range should be ${\bf 10^{-5}-10^{-3}}$ if the right-handed neutrino masses are in the order of TeV
and the ordinary neutrino masses are in the range of $10^{-3}-10^{-1}$ eV.
This is much larger than the estimation in Ref.\cite{1611.07975-neff_mth}
since it assumes that the right-handed neutrinos are in the GeV order.

\section{The processes $\ell_i\to \ell_j \gamma$ and $\ell_i \to \ell_j\ell_k\ell_l$ }

In this section, we analyze the branching ratios of certain LFV processes in the TH model, including $\mu \to e\gamma$, $\tau \to e(\mu)\gamma$, $\mu \to 3e$, $\tau \to 3e$, $\tau \to 3\mu$.
\subsection{The width and the Branching ratios of the rare decay $\ell_i\to \ell_j \gamma$}
\begin{figure}[H]
\centering
\includegraphics[width=10cm]{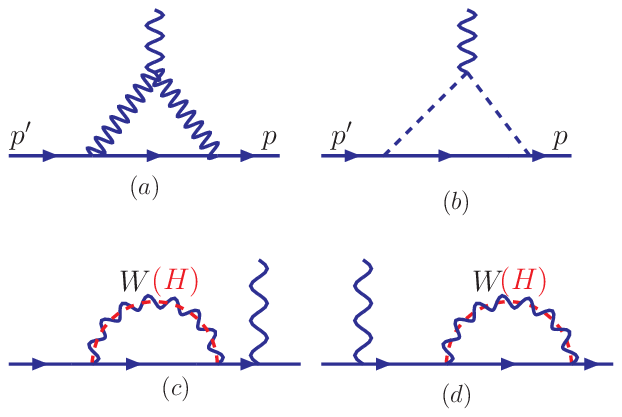} \vspace{-1.5cm}
\caption{ Fenyman diagrams for $\ell_i \to \ell_j \gamma$ decays.
The solid lines, wavy lines and dash lines denote the fermions, the gauge bosons and the charged Higgs, respectively,
which are the same as those in Fig.\ref{loop-3e}.}
\label{fig1}
\end{figure}
The Feynman diagrams for $\ell_i\to \ell_j \gamma$, which can be mediated by the gauge bosons and the charged Higgs, are shown in Fig.\ref{fig1}, with the initial and the final leptons $\ell_i$ and $\ell_j$, respectively.

The Lorentz structure of $\ell_i\to \ell_j \gamma$ is written as the matrix element of the electromagnetic current
between incoming and outgoing fermion states with momentum and spin $\{p,s\}$ and $\{p',s'\}$, respectively, 
\beq
<p,s|J_\lambda^{em}|p',s'>= \bar u_e(p,s)[F_1(Q^2)\gamma_\lambda +\frac{F_2(Q^2)}{2m}\sigma_{\lambda\nu} q^\nu ] u_p(p',s'),
\label{lor_llr}
\eeq
where $\sigma_{\lambda\nu}=\frac{i}{2}[\gamma_\lambda ,\gamma_\nu]$, and $q$ is the photon momentum, $q=p'-p$, $Q^2=-q^2$.
For the amplitude of $\ell_i\to \ell_j \gamma$,
 the first term in Eq.(\ref{lor_llr}) will disappear
due to the electromagnetic gauge invariance, $\partial^\lambda J_\lambda^{em}=0$.
 That can be seen as following.
 Multiplying Eq.(\ref{lor_llr}) by $\partial^\lambda=-iq^\lambda$, the left side is equal to zero.
For $Q^2\to 0$, the second term of the right side vanishes,
and thus the first term of the same side equals to zero, i.e, $F_1(0)=0$.
This term corresponds to the self energy diagrams in Fig.\ref{fig1}(c)(d) and
they do not contribute to the amplitude of the decay $\ell_i\to \ell_j \gamma$.

Therefore, the total effective amplitude of $\ell_i\to \ell_j \gamma$,
which is contributed by the gauge bosons and the charged Higgs $H^\pm$ with the heavy neutrinos,
 can be written as
\bea
{\cal M}_{total}&=&{\cal M}^{\nu W}+{\cal M}^{\nu H}  = m_i\cdot \bar u(p)\cdot M \cdot (1+\gamma^5)\cdot 2(p'\cdot \epsilon)] u(p'),
\eea
where $m_i(m_j)$ is the mass of the initial(final) state lepton,
 and  $p',~p,~p_\gamma$ are the momentums of the initial, final state lepton and the photon, respectively,
and $\epsilon$ is the polarization vector of the photon.
 Note that the final state lepton masses are neglected, $m_j=0$.
\bea
M&=&\frac{ ie }{32\pi^2}\{
  |V_{\tilde{\nu}}|^2 ( C_{12}-2C_{21}+2C_{23} )^{W}
+ |y_{\nu_R}|^2 ( C_{21}-C_{23}+C_{11} -C_{12} )^{H^\pm} \},
\label{mm}
\eea
where $V_{\tilde{\nu}}$ ($y_{\nu_R}$) is the couplings of the charged gauge bosons(Higgs) to
the fermions, as previously stated. Since gauge bosons $W^\pm$ couple to twin neutrinos, and charged Higgs $H^\pm$ to right-handed neutrinos,
so the terms with the "$W$" superscript are the contributions from twin neutrinos, and "$H$" superscript ones, from right-handed neutrinos.

In the above equation,  $C_{ij}^P=C_{ij}(p_\gamma,-p',(m^P_{\nu})^2,m_P^2,m_P^2)$ ($P=~H^\pm,~W$)
are the one-loop three point functions ~\cite{looptools}, and for $P=W^\pm$, $m^P_{\nu}=m_{\tilde\nu}$;
$P=H^\pm$, $m^P_{\nu}=m_{\nu_R}$.

So, the decay width is given as
\bea%
\Gamma(\ell_i\to \ell_j \gamma)=\frac{m^3_i}{4\pi}\cdot m_i^2 \cdot |M|^2 = \frac{m^5_i}{4\pi} \cdot |M|^2.
 \eea %

\subsubsection{The process $\mu\to e \gamma$}  

The width of the dominant muon-decay mode $\mu\to e \nu \bar\nu$ is given as,
\beq \Gamma^\mu=\Gamma(\mu\to e \nu \bar\nu)\simeq
m^5_{\mu}G^2_F/(192\pi^3)~,
\eeq
where $G_F=\frac{\sqrt{2}g^2}{8 m_W^2}$ is the Fermi coupling constant. 
So the branching ratios of the rare decay $\mu\to e \gamma$ is given by
\begin{eqnarray}
BR(\mu\to e \gamma) = \frac{\Gamma(\mu\to e
\gamma)}{\Gamma(\mu\to e \nu \bar\nu)}
=\frac{48 \pi^2 }{G_F^2} \cdot |M|^2,
\end{eqnarray}
%
\begin{figure}[H]
\centering
\includegraphics[width=6.5cm]{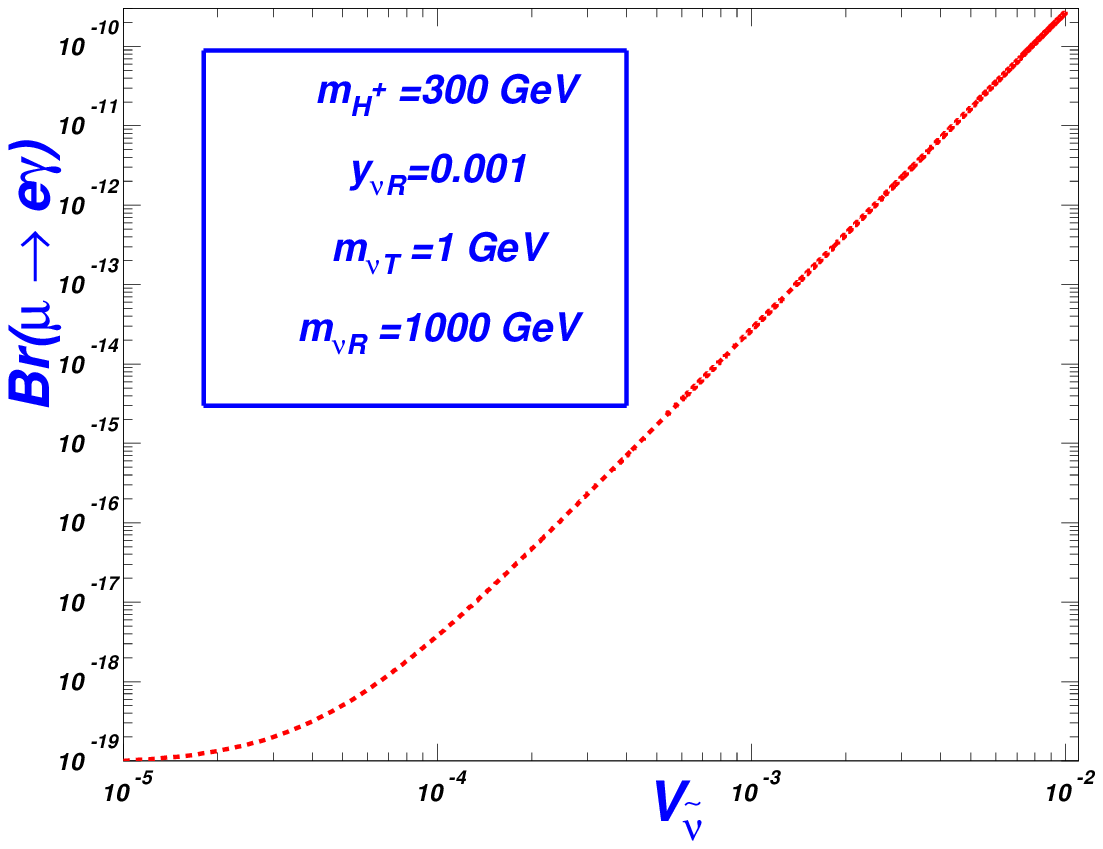}  
\includegraphics[width=6.5cm]{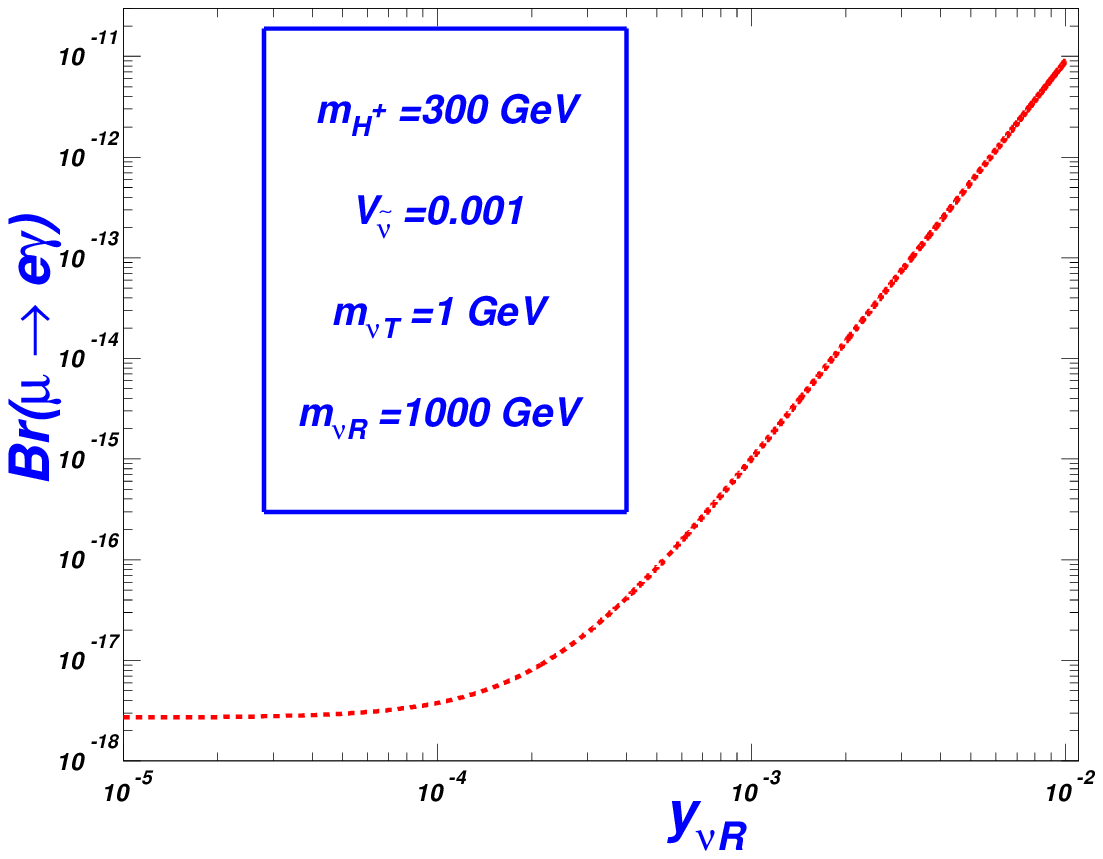}\\
\includegraphics[width=5cm]{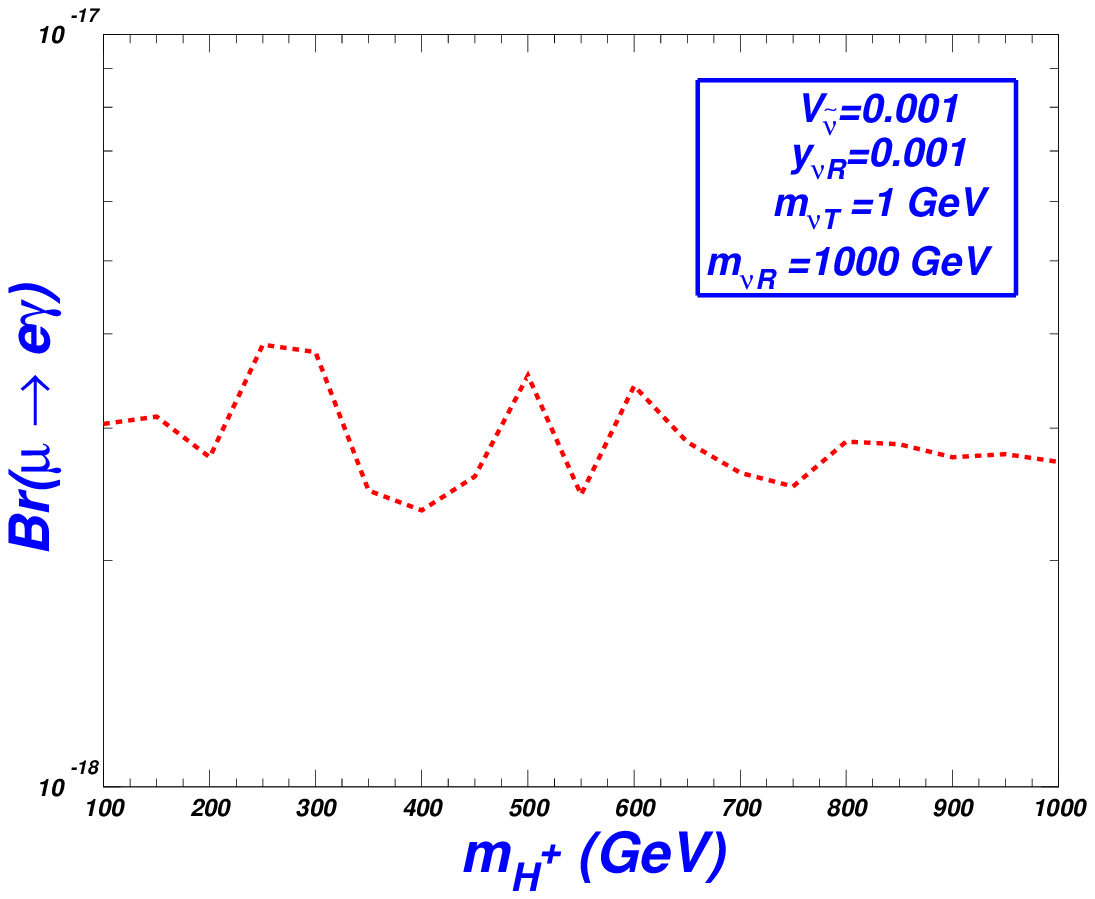}  
\includegraphics[width=5cm]{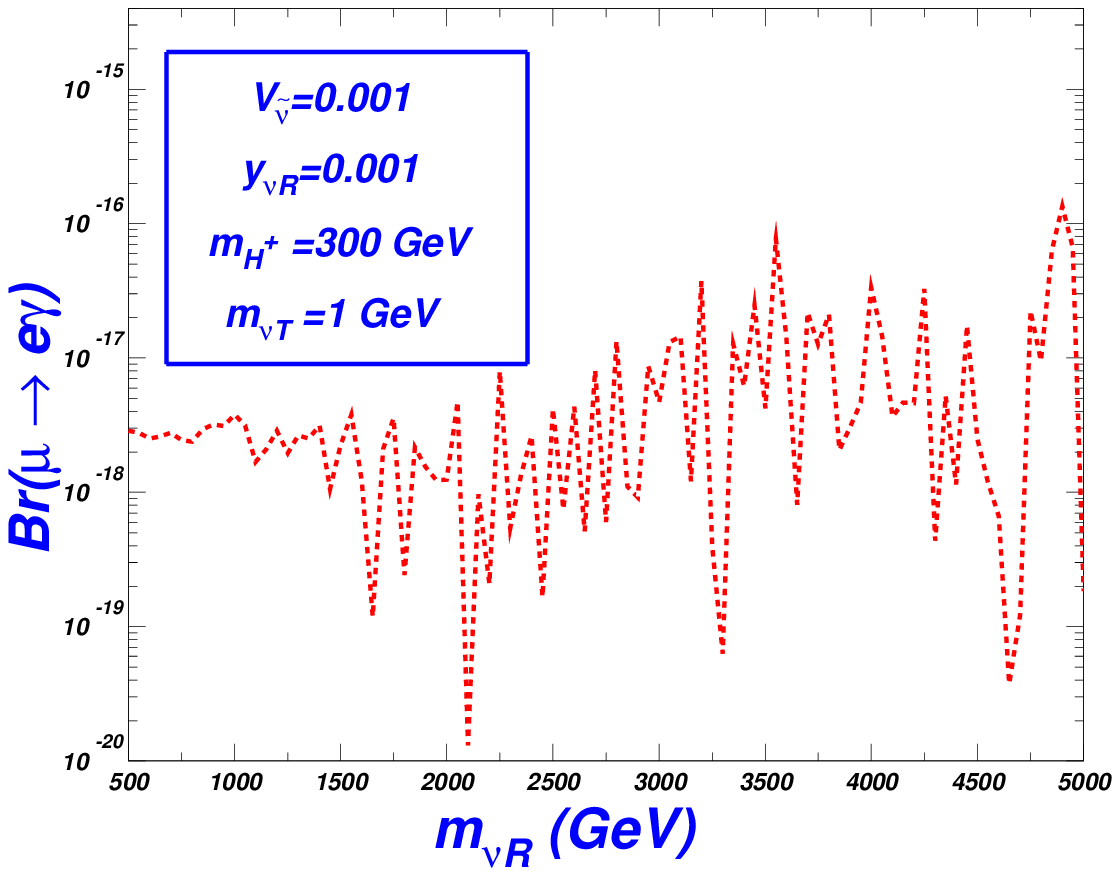} 
\includegraphics[width=5cm]{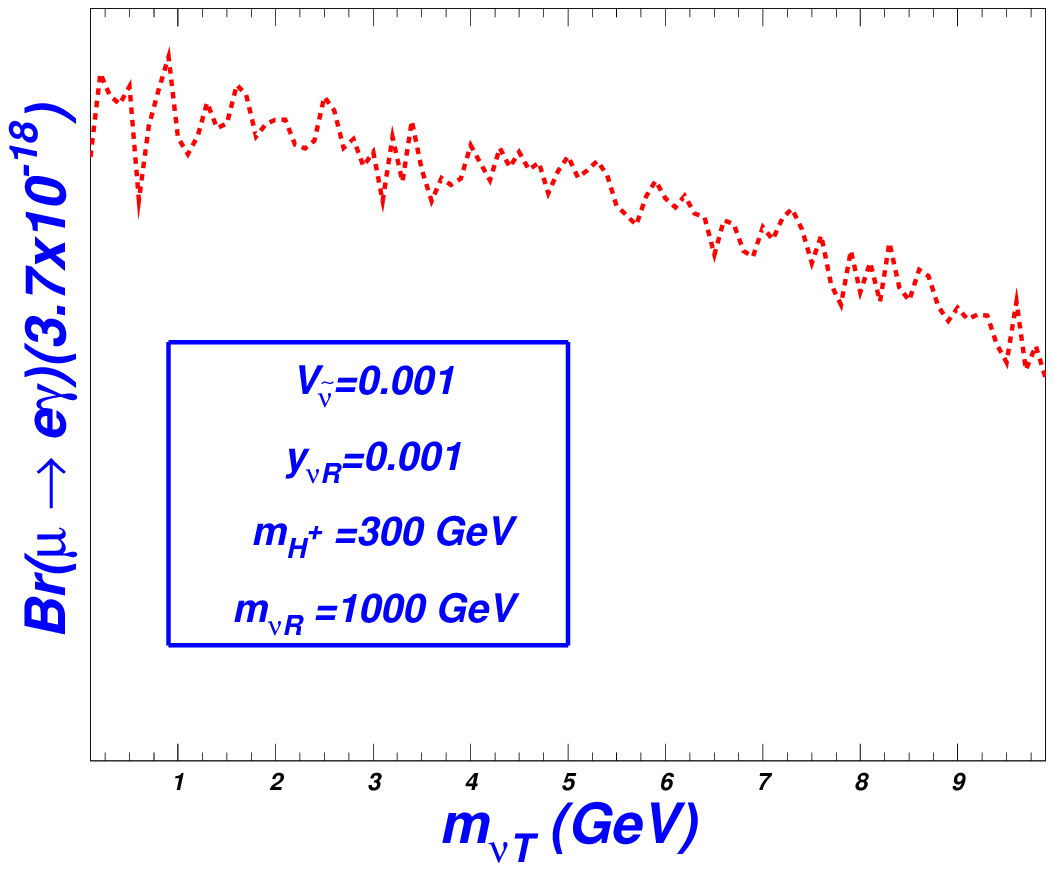}
\caption{With other parameters shown in the figures, the branching ratios of $\mu\to e \gamma$ 
versus parameters $V_{\tilde \nu},~y_{\nu_R},~m_{H^\pm},~m_{\tilde \nu},~m_{\nu_R}$, respectively.} \label{fig2}
\end{figure}

 For the third figure of Fig.\ref{fig2}, the value range of Y-axis is $(2.7-3.1) \times 10^{-18}$, in the same order,
 so we let it vary between $10^{-18}$ to $10^{-17}$.
 For the last figure of Fig.2, the number $3.7\times 10^{-18}$ labeled next to the Y-axis variable $BR(\mu\to e\gamma)$
 is the middle value of the narrow range of the Y-axis:($3.6-3.8$)$\times 10^{-18}$.

In the lower part of  Fig.\ref{fig2}, we can see the contributions
to the branching ratios from the masses of the charged Higgs and the neutrinos
are small and  oscillatory in the narrow ranges.\footnote{The new particles,
 such as charged Higgs, right-handed neutrinos and the heavy fermions, are in the inner lines of the Fig. \ref{fig1},
 and the contributions are expressed by complex integral function $C_{ij}$.
According to the definition of the integral function $C_{ij}$, there are terms such as $\frac{1}{p^2-m^2+i\epsilon}$
in the denominator($\epsilon$ is a very tiny real number),
and in general calculation, the term $i\epsilon$ is neglected,
but it will play a crucial role when the momentums are close to the masses, ensuring the results converging.
This situation may lead to oscillating results in some mass range.}
\footnote{Generally speaking, the contribution decreases with the increasing masses.
 In Fig.\ref{fig2}, since the mass ranges are not large enough, this decreasing effect is not obvious.
 But it is impossible for the contributions to oscillate regularly to very large masses.
 Even so, with very narrow varying range, from the last figure of Fig.\ref{fig2},
 we can still see the decreasing branching ratio with the increasing twin neutrino mass.}
Moreover, they are all smaller than the detected level given in Eq.(\ref{mubound}).
Thus in the following, we will take the masses as fixed values.

In the upper part of Fig.\ref{fig2}, we show the dependence of the $\mu\to e \gamma$ branching ratios on the couplings,
$V_{\tilde \nu},~y_{\nu_R}$,
from which we can see that the branching ratios increase monotonously along with the increasing couplings,
and that they may arrive at the detectable level,
so the couplings play a more important role on the branching ratios in comparison with the mass parameters.

The $\mu\to e \gamma$  detectable experimental upper bound is given by~\cite{1605.05081,1107.5547}
\beq
BR(\mu \to e \gamma) < 4.2\times  10^{-13}.
\label{mubound}
\eeq
From this constraints, the approximate range of the couplings can be given as
 \beq
 V_{\tilde \nu}< 0.00198, ~~ y_{\nu_R}< 0.0046.
 \label{constr-coup}
 \eeq
But if it is possible for the processes to be detected in experiments, the couplings
$V_{\tilde \nu},~y_{\nu_R}$  should be larger than $0.00198, ~ 0.0046$, respectively,
which would be stringent constraints.

\begin{figure}[H]
\centering
\includegraphics[width=8cm]{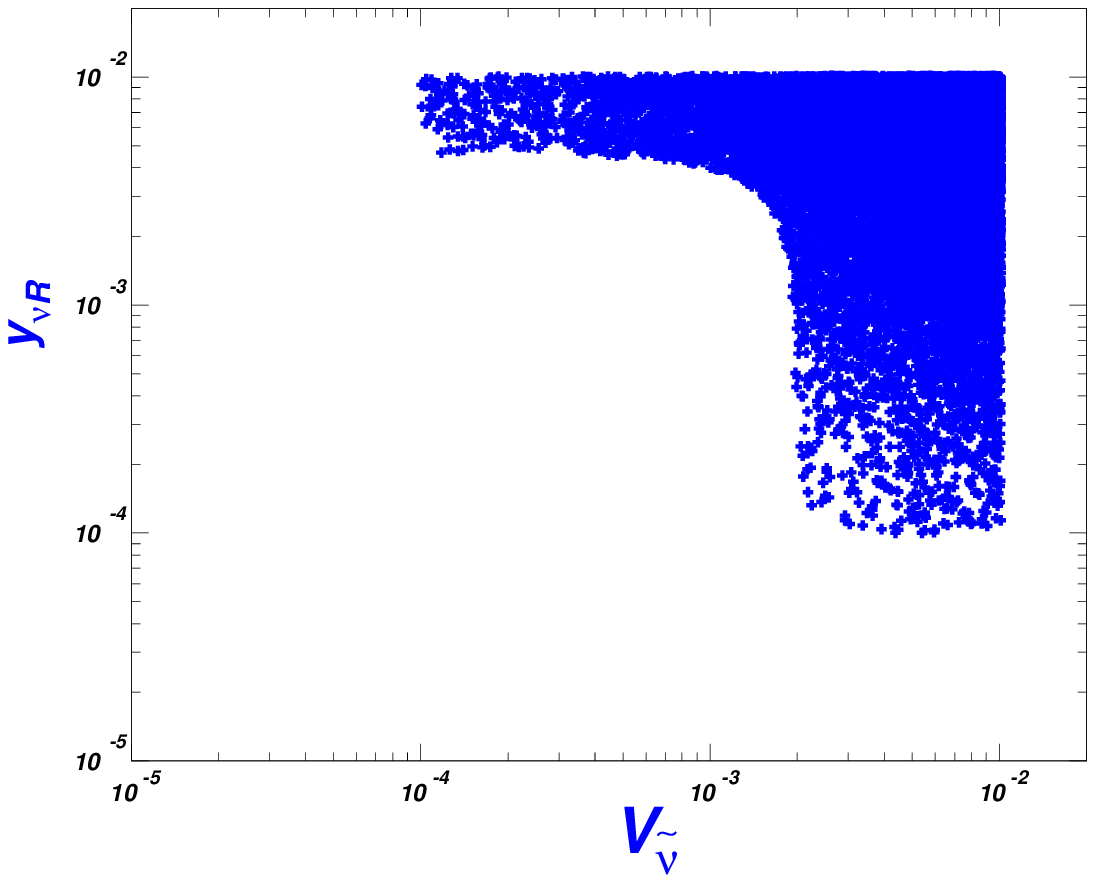}\\
\caption{With $m_{\tilde \nu}=1~GeV,~m_{\nu_R}=1000~GeV,~m_{H^\pm}=300~GeV$,
the allowed points of $V_{\tilde \nu}$  and $y_{\nu_R}$
to be larger than the experimental upper bound for the $\mu\to e \gamma$ branching ratios.} \label{fig3}
\end{figure}

The above constraints are derived from the independent influence of these two couplings on the branching ratio.
We can also consider the points allowed to exist in the scan parameter space
when the branching ratios are greater than the upper limit of the experimental upper bound.
%
We scan the coupling parameters in the ranges:($10^{-5}-10^{-2}$)
and their possible allowed points are shown in Fig.\ref{fig3},
from which we can see that, for this process, large $V_{\tilde \nu}$ or $y_{\nu_R}$ is favorable to be detected at the colliders.

\subsubsection{The processes  $\tau \to e\gamma$ and  $\tau \to \mu\gamma$}


As for the $\tau \to e\gamma$ and  $\tau \to \mu\gamma$, the total widths are different from that of the muon
since the $\tau$ can decay not only into $e\nu\bar \nu$, but also into $\mu\nu\bar{\nu}$.
Besides, it can decay hadronically since lepton $\tau$ is heavier than the light quarks.
Taking into account the unitarity of the quark mixing matrix, $|V_{ud}|^2 + |V_{us}|^2 = 1 -|V_{ub}|^2 \simeq 1$, one can easily get the lowest-order width (see e.g. Ref.~\cite{1310.7922} and the Refs. in it),
\beq
\Gamma_{\tau; total}=\Gamma_\mu \frac{m_\tau^5}{m_\mu^5} [2+N_C( |V_{ud}|^2+|V_{us}|^2)] \simeq 5 \cdot \Gamma_\mu \frac{m_\tau^5}{m_\mu^5},
\label{wid_tau}
\eeq
so the branching ratios of $\tau \to e\gamma$ and  $\tau \to \mu\gamma$ can be written as,
\begin{eqnarray}
BR(\tau\to e/\mu \gamma) = \frac{\Gamma(\tau\to e/\mu \gamma)}{\Gamma^\tau_{total}}
=\frac{48 \pi^2 }{5G_F^2} \cdot |M|^2,
\label{tau_br}
\end{eqnarray}
Note that the notation $M$ is the same as Eq.(\ref{mm})
except that the leptons involved in their coupling coefficients are different: The former is $\mu$ with $e$, and the latter is $\tau$ with $e$ or $\mu$,
but the expressions are the same since there will be no confusion with each other.


The experimental upper bounds are ~\cite{taubound-0908.2381},
\beq
BR(\tau\to e \gamma)< 3.3 \times 10^{-8},~~~~~~BR(\tau\to\mu\gamma)< 4.4 \times 10^{-8}.
\label{taubound}
\eeq

The two LFV processes $\tau\to e \gamma$ and $\tau\to \mu \gamma$,
are almost the same. What makes difference is the final lepton mass,
but in the crude estimation, we can neglect the final lepton masses, i.e, $m_e=m_\mu=0$.
Therefore, in the following we will just consider the former.
%
%
\begin{figure}[H]
\centering
\includegraphics[width=8cm]{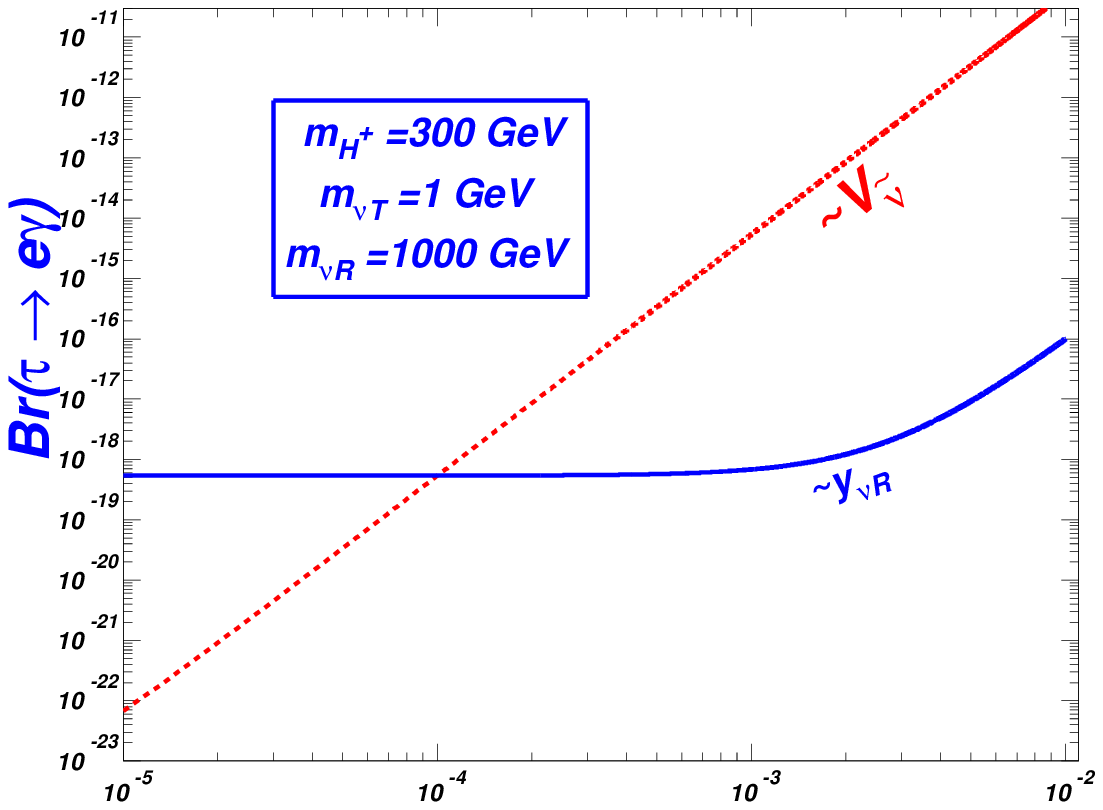}\\
\caption{With $m_{\tilde \nu}=1~GeV,~m_{\nu_R}=1000~GeV,~m_{H^\pm}=300~GeV$,
the branching ratios of $\tau\to e \gamma$ 
versus parameters $V_{\tilde \nu}$ (for $y_{\nu_R} =0.0001$)
and $y_{\nu_R}$ (for $V_{\tilde \nu} =0.0001$), respectively.} \label{fig4}
\end{figure}

In Fig.\ref{fig4}, we show the branching ratios of $\tau\to e \gamma$ vary with
the couplings $V_{\tilde \nu}$ and $y_{\nu_R}$ with the fixed masses of the twin neutrino, right-handed neutrino and the charged Higgs,
since they are not sensitive to the varying of the mass parameters.
We take  $y_{\nu_R} =0.0001$ ($V_{\tilde \nu} =0.0001$) when the variable is $V_{\tilde \nu}$ ($y_{\nu_R}$).
Fig.\ref{fig4} shows that the branching ratios depend greatly on the couplings.
However, even when $V_{\tilde \nu}$ or $y_{\nu_R}$ is close to $0.01 $, which is quite large for the ranges discussed in Sec. II,
the branching ratios can still not reach the detectable sensitivity
in experiments shown in Eq.(\ref{taubound}).
Thus this process can not provide constraints to the couplings for the time being,
so we will not discuss it further.



\subsection{The decay $\mu\to 3e$ and $\tau\to \ell e^+e^-$ ($\ell=e,~\mu$)}

The current 90\% C.L. bounds on the decay modes of three bodies are ~\cite{pdg-2018,tau23e},
\beq
BR(\mu \to 3 e) < 1.0\times 10^{-12}, \\   BR(\tau\to 3e(e\mu\bar\mu)) < 2.7\times 10^{-8},
 BR(\tau\to 3\mu ) < 2.1 \times 10^{-8}.
 \label{3l-bounds}
\eeq

\subsubsection{The decay $\mu\to 3e$ and $\tau\to 3e$ mediated by new neutral gauge bosons} 

We now discuss the three body decays of the leptons mediated by the new neutral gauge bosons. 
The so-called new neutral gauge bosons mean the partners of the SM gauge bosons or the
extra vector gauge boson introduced by the twin Higgs modes, such as Ref. \cite{1811.05977},
which adds a singleton vector $X_\mu$ to link the SM and the twin gauge sector.
Their couplings to the leptons can be written in a general form as
\beq
{\cal L}  \sim  V_\mu \gamma^\mu\bar\ell(g^{(V)}_{V\ell\ell'}+g^{(A)}_{V\ell\ell'}\gamma^5) \ell'
+ V_\mu \gamma^\mu\bar\ell(g^{(V)}_{V\ell\ell}+g^{(A)}_{V\ell\ell}\gamma^5) \ell +h.c.,
\eeq
where the gauge boson is denoted as $V_\mu$,
and the superscript $(V)$ and  $(A)$ denote the parts of vector and axis vector of
the boson couplings to the leptons. 
 We will check the contribution from the new gauge bosons to simply estimate the parameter limits.

On the other hand, an important reason to consider the gauge boson effect
on the tree-level decay process $\ell_i\to \ell_j\ell_j\ell_j$
is that the contribution mediated by the neutral Higgs scalar
is much smaller than that of the neutral gauge boson,
since the lepton Yukawa couplings i.e, $m_\tau/v,~m_\mu/v $, should be much smaller than the gauge coupling.

Since neutral gauge boson is much heavier than muon,
the width of the three-body decay $\mu \to 3e$ can be written as
\bea\label{eq:BrMu3E}  \nm
{\rm BR}(\mu^{-}\to e^{-} e^{+}e^{-}) &=&
\frac{\Gamma(\mu^{-}\to e^- e^{+}e^{-})}{\Gamma(\mu^{-}\to e^- \nu\nu )} \\
&=& \frac{\kappa\tilde{g}_{ \mu e}^2
\tilde{g}_{ee}^2/M_V^4}{\kappa g_W^4/M_W^4}
 = \frac{\tilde{g}_{ \mu e}^2
\tilde{g}_{ee}^2}{ g_W^4}
\frac{M_W^4}{M_V^4},
\eea
where $\tilde{g}_{\mu e}^{2} = |g^{(V)}_{V\mu e}|^{2}+|g^{(A)}_{V\mu e}|^{2}$,
 $\tilde{g}_{e e}^{2} = |g^{(V)}_{Ve e}|^{2}+|g^{(A)}_{Ve e}|^{2}$,
$M_V$ is the mass of the gauge boson, $\kappa$ 
is a kinematic-spin factor common to all decay modes mediated by the
 neutral vector boson, while
$g_{W}$( $G_F/\sqrt{2} =  g_W^2 /(2 M_W^2)$) and $M_{W}$ are the electroweak coupling and the $W$ boson mass, respectively.

 Thus via the same idea, the three-body decay of the lepton $\tau$ involving
vector gauge bosons in the intermediate state can be written as
\begin{eqnarray}\label{eq:TauLEE-1} \nm
{\rm BR}(\tau^{-}\to \ell^{-} e^{+}e^{-}) &=&
\frac{\Gamma(\tau^{-}\to \ell^{-} e^{+}e^{-})}{\Gamma(\tau^{-}\to \ell^{-} \nu\nu )}
\frac{\Gamma(\tau^{-}\to \ell^{-} \nu\nu )}{\Gamma(\tau^-\to All)} \\ \nm &=&
\frac{\Gamma(\tau^{-}\to \ell^{-} e^{+}e^{-})}{\Gamma(\tau^{-}\to \ell^{-} \nu\nu )}
\frac{\Gamma(\tau^{-}\to \ell^{-} \nu\nu )}{\Gamma(\mu^-\to  e^-\nu\nu)}
\frac{\Gamma(\mu^-\to  e^- \nu \nu)}{\Gamma(\tau^-\to All)}
\\
&=&  \frac{\tilde{g}_{ \tau\ell}^2
\tilde{g}_{ee}^2}{g_W^4}
\frac{M_W^4}{M_V^4}
\left(\frac{M_{\tau}}{M_{\mu}}\right)^5
 \frac{\Gamma^\mu_{total}}{\Gamma^\tau_{total}}
\end{eqnarray}
Note that in the above formulas, the final lepton masses have been neglected.

\begin{figure}[H]
\centering
\includegraphics[width=8cm]{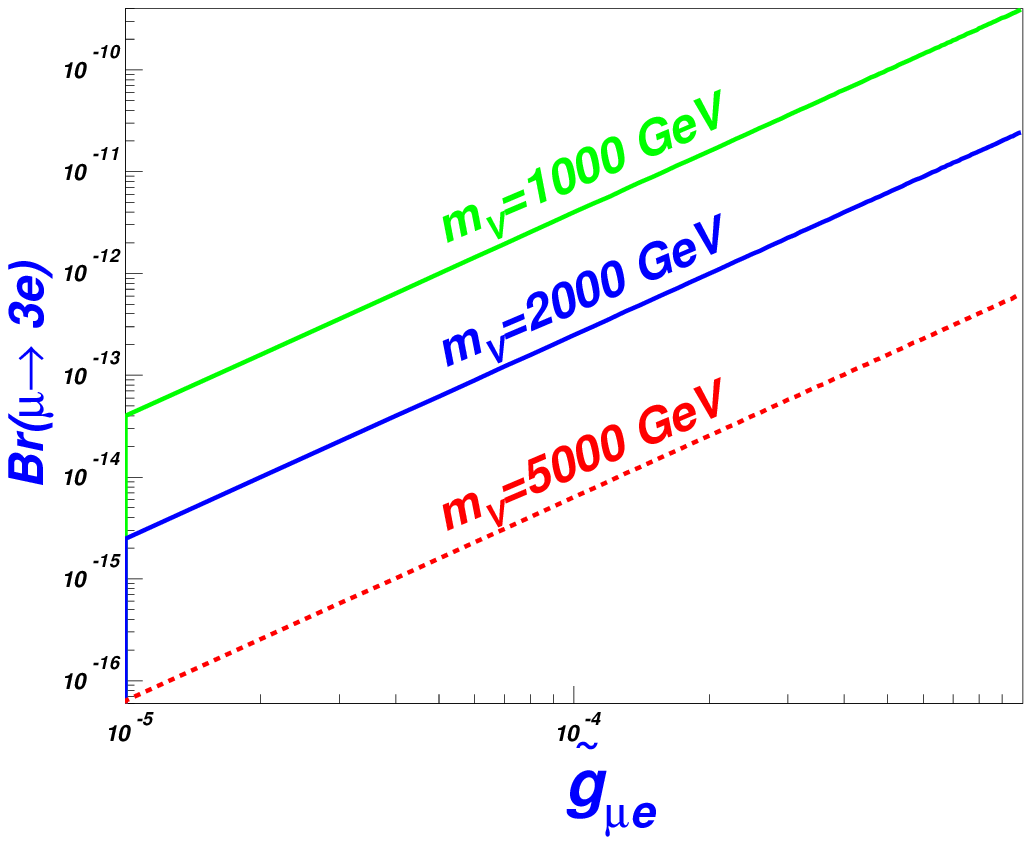}
\includegraphics[width=8cm]{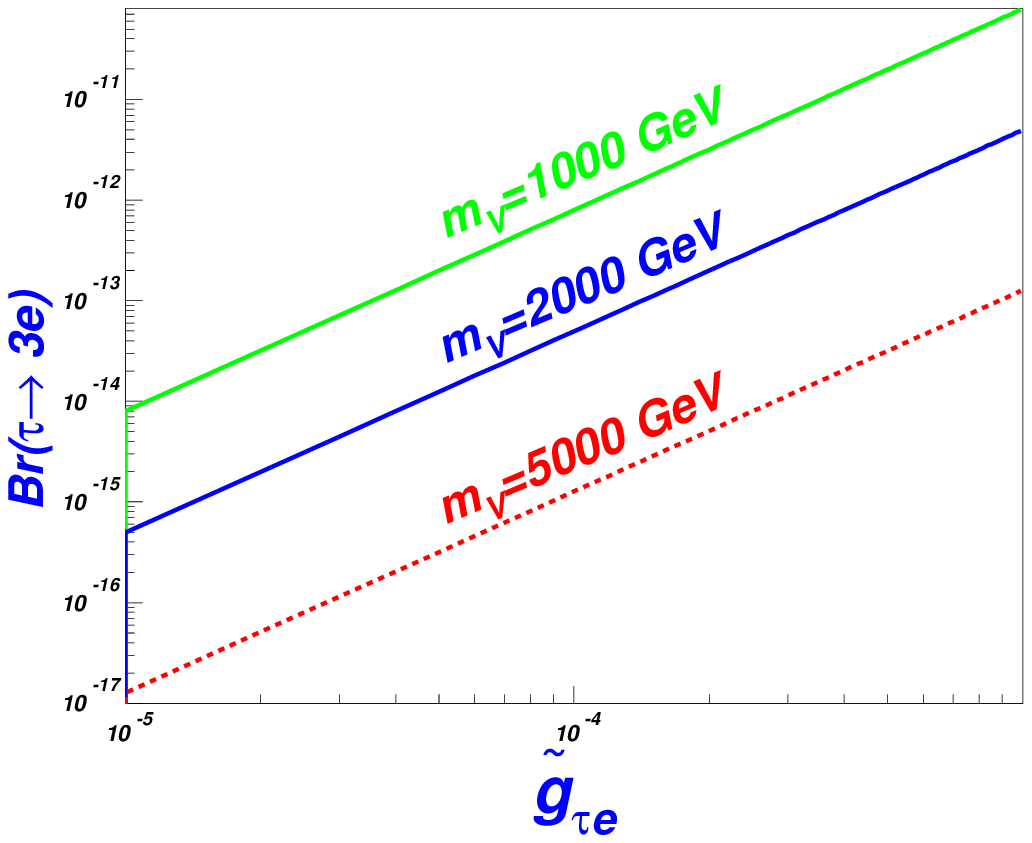}\\
\caption{The branching ratios of $\mu\to 3e $(left) and $\tau\to 3e $(right) versus parameters $\tilde{g}$
for $m_{Z_0^\pr}=1000,~2000,~5000$ GeV, respectively.} \label{fig5}
\end{figure}

The dependence of the branching ratios on the parameters $\tilde{g}$ and the mediator gauge boson mass $m_V$
are given in Fig.\ref{fig5}  for
$\mu\to 3e$ and $\tau\to 3e $.
We find that the branching ratios can account for the the experimental bounds Eq.(\ref{3l-bounds})
in quite a large parameter space with a not too large extra new neutral gauge boson,
especially for decay $\mu\to 3e$.

According to the constraints in Eq.(\ref{3l-bounds}),
to probe the processes, the lower bounds of the couplings $\tilde{g}_{\mu e}, \tilde{g}_{\tau e}$
can be estimated as
\beq
\tilde{g}_{\mu e} \geq 0.000051, ~~ \tilde{g}_{\tau e} \geq 0.018~,
\label{mue-taue-bound}
\eeq
with $M_V =1000~$GeV.
The main reason of the bound on $\tilde{g}_{\mu e}$ is stronger than $\tilde{g}_{\tau e}$
in Eq.(\ref{mue-taue-bound}) for a fixed $M_V$ is that, according to Eq.(\ref{3l-bounds}),
to be detected in the experiments,
the needed branching ratio of $\mu\to 3e$ should be four orders smaller than that of $\tau\to 3e$,
while their branching ratios are almost in the same order, just as shown in Fig.\ref{fig5}.
And from Fig.\ref{fig5} we can also see that the branching ratios increase with the increasing couplings.
Thus $\tilde{g}_{\mu e}$ can be much smaller than $\tilde{g}_{\tau e}$ for the processes to be probed.

At the same time, the branching ratio of the former is about five times that of the latter.
So the bound on $\tilde{g}_{\mu e}$ is much stronger,
which would be a good feature to probe the new gauge boson
via the decay $\mu\to 3e$ since in many models, a small coupling is much more viable.

\subsubsection{The decay $\mu\to 3e$ and $\tau\to 3e$ mediated by the charged scalar Higgs and the right handed neutrino at the one-loop level}

Since there are no tree-level contributions
 to the $\ell_i\to \ell_j\ell_j\ell_j$ decay modes by the charged bosons and Higgs,
 some works, such as Refs.~\cite{1702.04399,1909.12147,1909.02044}, 
consider these processes at the one-loop level.
For completion, we will refer to the above references to check the contributions from
 the charged Higgs and the charged gauge bosons with the extra heavy neutrino.

\begin{figure}[H]
\centering
\includegraphics[width=15cm]{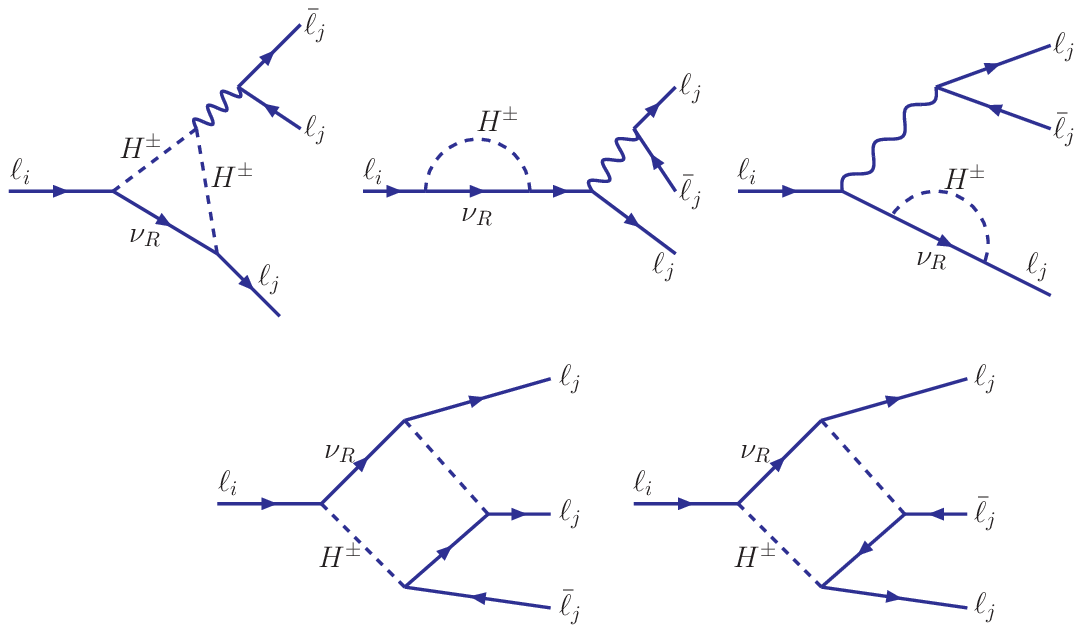}\\
\caption{ The one-loop charged scalar contributions of the LFV couplings to the three final
leptonic states of the second or the third generation lepton.
} \label{loop-3e}
\end{figure}
The couplings between the charged Higgs, the charged gauge bosons with the extra heavy neutrinos and the charged leptons can induce the three-body decay of the massive charged leptons at one-loop level.
Typical Feynman diagrams are shown in Fig.\ref{loop-3e} and
the branching ratio can be
found in the Appendix.

\begin{figure}[H]
\centering
\includegraphics[width=9cm]{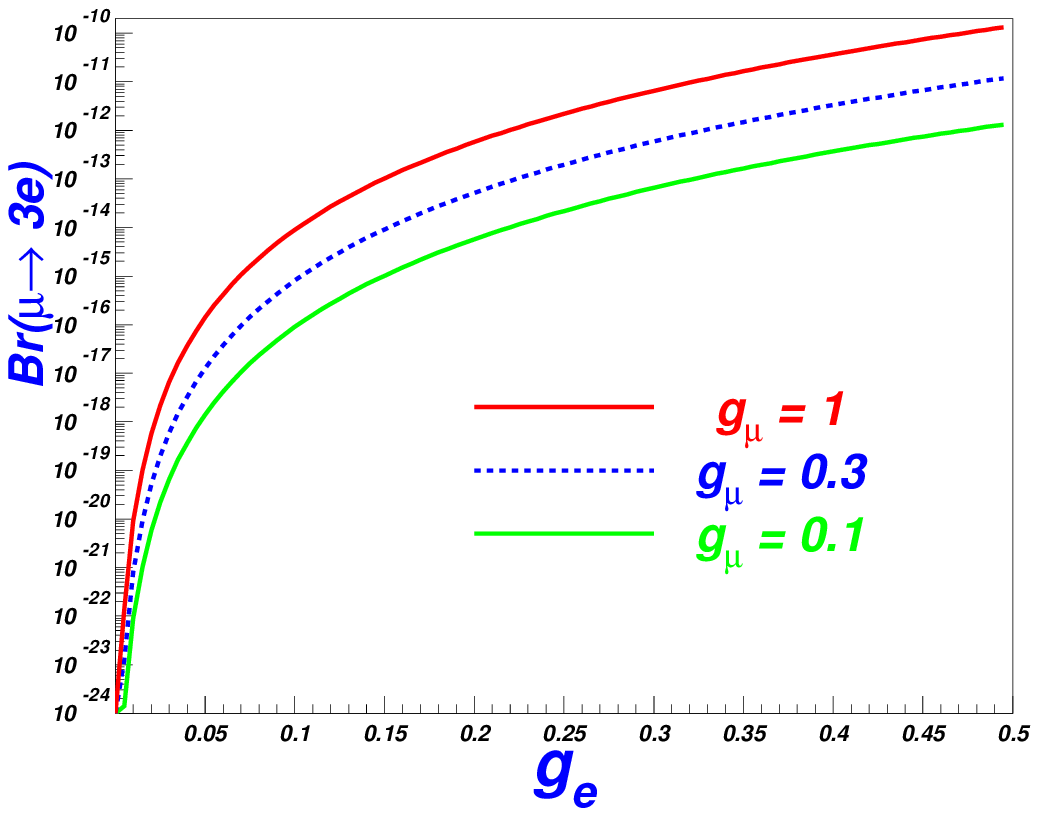}\\
\caption{The branching ratios of $\mu\to 3e $ at the one-loop level versus parameters $g_e$ with
 $g_\mu =0.1,~0.3,~1$, respectively.} \label{fig12}
\end{figure}

With fixed masses for the charged scalar Higgs and the heavy neutrinos, the branching ratios
at the one-loop level are shown in Fig.\ref{fig12},
in which $g_{\mu}$ ($g_{e}$) is the coupling of the charged Higgs to the lepton $\mu$ ($e$) and the right-handed neutrino,
similar to $y_{\nu_R}$ in Eq.(\ref{lfv-2}), but now distinguished for different flavors.
We can see that the decay rates are still below the detection sensitivity in Eq.(\ref{3l-bounds}) even if we choose the couplings to be ${\cal O}(1)$.
Similar discussions can be given for the same three-body decays mediated by the gauge bosons and the heavy neutrinos at one-loop level.
We, however, neglect such processes since their contributions are too small compared to the experimental bounds.

\section{Conclusion}
We discuss the possible effects of the new exotic fields $H^\pm$ together with the right-handed neutrinos
and gauge bosons $W^\pm$ with the twin neutrinos in the TH models
on the LFV processes $\ell_i \to \ell_j \gamma$ and $\ell_i \to \ell_j \ell_k \ell_l$, respectively.

From the discussion given above, with the assumption of degenerate neutrinos for simplification,
we can draw the following conclusions:
\begin{itemize}

\item In some parameter space such as $V_{\tilde \nu}> 0.00198$ or $y_{\nu_R}> 0.0046$,
the decay channel $\mu \to e \gamma$ may arrive at the detectable level.
On the other hand, if the experiments can not find any signal of this channel,
since the decay $\mu \to e \gamma$ is sensitive to the couplings $\tilde V_\nu$ and $y_{\nu_R}$,
this may provide constraints on the parameters in models of this kind.


\item Similarly, the three body decay channel $\mu \to 3e$ mediated by the neutral gauge boson may arrive at the
the detection level of the experiments,
while those mediated via the charged Higgs and the gauge bosons at the one-loop level can not be possible
to be detected in most parameter space.

\item But this is not true for $\tau \to e \gamma$ and $\tau \to 3e$ decay channels.
They cannot be probed in the whole possible parameter space.  
\end{itemize}

Therefore, we conclude that the LFV processes of TH models may arrive at the detectable level in a large parameter space,
and they may shed light on the constraints of the models or even to find some signal of the models.

\vspace{1cm}

{\bf Acknowledgments}

 This work was supported by the National Natural Science Foundation of China(NSFC)
under grant 11675147,12075213, 
11775012, 
 by the Key Project by the Education Department of Henan Province under grant number 21A140025,
 by the Fundamental Research Cultivation Fund for Young Teachers of Zhengzhou University(JC202041040)
 and the Academic Improvement Project of Zhengzhou University.

\appendix{}
\section{}
The branching ratio of $\mu \to 3 e $ mediated by charged Higgs can be written as\cite{1702.04399,1909.12147}
%
%
\begin{align}
BR(\ell_{i}\rightarrow\ell_{j}\bar{\ell}_{j}\ell_{j}) & = \frac{3 e^{2}}{8G_{F}^{2}}\Bigg[|A_{ND}|^{2}+|A_{D}|^{2}\left(\frac{16}{3}\log\left(\frac{m_{i}}{m_{j}}\right)-\frac{22}{3}\right)\nonumber\\
& +\frac{1}{6}|B|^2 +\frac{1}{3}\frac{m_{i}^{2}m_{j}^{2}\left(3\sin^{4}\theta_{W}-\sin^{2}\theta_{W}+\frac{1}{4}\right)}{m_{W}^{4}\sin^{4}\theta_{W}}\left|A_{D}\right|^{2}\nonumber\\
& +\left(-2A_{ND}A_{D}^{\ast}+\frac{1}{3}A_{ND}B^{\ast}-\frac{2}{3}A_{D}B^{\ast}+\mathrm{h.c.}\right)\Bigg] \nonumber\\
& \times BR(\ell_{i}\rightarrow\ell_{j}\nu_{i}\bar{\nu}_{j}),\label{tmug}
\end{align}
where 
$A_D$ is the dipole contribution,
\begin{equation}
A_D= \frac{g_{j}^{\ast}g_{i}F(x)}{2(4\pi)^2 m^2_{H^\pm}},
\end{equation}
where $g_{i}$ ($g_{j}$) ($i,j=e,~\mu,~\tau$) is the couplings of the charged Higgs to the lepton $i$ ($j$) and the right-handed neutrino,
$m_{H^\pm}$ the mass of heavy Higgs ($H^\pm$)
 and $x=\frac{m^2_{vR}}{m^2_{H^\pm}}$. Expression of $F(x)$ will be given shortly. 
The coefficients $A_{ND}$ and $B$ are the non-dipole contributions
from the penguin and the box diagrams, respectively, which read
\begin{equation}
A_{ND}=\frac{g_{j}^{\ast}g_{i}}{6(4\pi)^{2}}\frac{1}{m_{H^\pm}^{2}}G\left(x\right),\label{AND}
\end{equation}
and
\bea
B 
&=&\frac{1}{(4\pi)^{2}e^{2}m_{H^\pm}^{2}}
\left[\frac{1}{2}D_{1}(x)g_{j}^{\ast}g_{j}g_{j}^{\ast}g_{i}
+x~D_{2}(x)g_{j}^{\ast}g_{j}^{\ast}g_{j}g_{i}\right].
\label{B}
\eea
Note that only one flavor of (mediated) neutrino is assumed.
\begin{align}
F\left(x\right) & =\frac{1-6x+3x^{2}+2x^{3}-6x^{2}\log x}{6\left(1-x\right)^{4}},\\
G\left(x\right)= & \frac{2-9x+18x^{2}-11x^{3}+6x^{3}\log x}{6\left(1-x\right)^{4}},\\
D_{1}(x) & =\frac{-1+x^{2}-2x\log x}{\left(1-x\right)^{3}},\\
D_{2}(x)= & \frac{-2+2x-\left(1+x\right)\log x}{\left(1-x\right)^{3}},
\end{align}
and
\begin{equation}
F\left(1\right)=\frac{1}{10},\qquad G\left(1\right)=\frac{1}{4},\qquad D_{1}(1)=\frac{1}{3},\qquad D_{2}\left(1\right)=\frac{1}{6}.
\end{equation}

 \end{document}